 \documentclass[final,5p,times,twocolumn]{elsarticle}

\usepackage{amssymb}

\usepackage{lineno}

\usepackage{lipsum}

\newcommand\ceil[1]{\lceil#1\rceil}

\journal{Journal Name}

\begin{document}

\begin{frontmatter}



\title{A Novel Approach to Implement Message Level Security in RESTful Web Services}


\author[label1]{Gyan Prakash Tiwary}
\author[label2]{Abhishek Srivastava}

\address{Discipline of Computer Science and Engineering\\
Indian Institute of Technology Indore\\
Indore, India 452020\\}
\address[label1]{Email: phd1501101003@iiti.ac}
\address[label2]{Email: asrivastava@iiti.ac.in}

\begin{abstract}
The world is rapidly adopting RESTful web services for most of its tasks. The once popular  SOAP-based web services are fast losing ground owing to this. RESTful web services are light weight services without strict message formats. RESTful web services, unlike SOAP, are capable of message transfer in any format be it XML, JSON, plain-text. However, in spite of these positives, ensuring message level security in REST is a challenge. Security in RESTful web services is still  largely dependent upon transport layer security. There has been some work recently towards message level security in such environments wherein the transfer of message level security metadata is done through utilising new HTTP headers. We feel, however, that any method that compromises the generality of the HTTP protocol should be avoided. In this paper, therefore, we propose two new ways of encryption that promise to ensure message level security in RESTful web services without the need for special HTTP headers. This approach works seamlessly on most famous content-types of RESTful web services: XML, JSON, HTML, plain-text and various ASCII printable content types. Further, the proposed approach removes the need for content negotiation in cases where the content comprises XML, JSON, HTML, plain-text, and ASCII printable content types and also removes the need for XML or JSON canonicalization. 
\end{abstract}

\begin{keyword}
RESTful Web Services \sep Message level Security


\end{keyword}

\end{frontmatter}


\section{Introduction}
\label{S:1}
Web services are a means to access the web in a `programmatic' manner. A website and a web service are similar in that both respond to requests made by clients on the web. Websites respond with content that can easily be comprehended by a  human eg. HTML, CSS and Javascript,  whereas web services respond with  content meant for consumption by other applications eg. XML, JSON. The response from web services has a greater focus on data whereas that from websites is more towards an interactive representation of data.  

\par  Web services may be categorised into two broad types: SOAP-based web services and RESTful web services. SOAP-based web services deal with properly structured messages comprising formal XML-based formats for a request and response. RESTful web services, on the other hand, lack such formal formats. This contributes to the flexibility and `light weight' nature of RESTful web services. SOAP request and response messages usually utilise HTTP or SMTP packets as containers, whereas REST requests comprise a simple HTTP request with the use of the common HTTP verbs CRUD (create, read, update and delete) for executing operations on the resource [1]. The response from a RESTful web service to a request is also an HTTP response containing XML, JSON, CSV, HTML, or plain-text.  

\par In REST, any content or service on the web for which a client makes a request to a server is known as a `resource'. A resource is content available with the server side and may get transferred or even modified based on a client's request. A resource may be a text or a binary file or may be data stored in a database. Each resource is identified by a unique ID called the Uniform Resource Identifier (URI). A resource may be represented in various formats such as XML, JSON, CVS, HTML, plain-text and others with the most popular representation today being XML and JSON.

\par RESTful web services, unlike SOAP, do not have a formal description language, therefore, it is common for services to publish their resource representations on public domains like websites. For example, Google publishes the resource representation of its drive API at the following URL \textit{https://developers.google.com/drive/v3/reference/about\# resource}. In addition to this, web service providers also describe the capabilities of the various HTTP verbs for performing operations on a given resource in a similar manner.

\par An important requirement for web services while connecting with their clients is a robust security mechanism. This is along lines similar to ordinary websites. Security is imperative for the following three purposes: confidentiality, integrity and authenticity. Confidentiality implies that a conversation between the server and the client makes no sense to a potential intruder eavesdropping on the conversation. Confidentiality can be provided by encryption of the message. Integrity implies that a message in transit between a client and a server must remain unchanged during transit.  Authenticity ensures that a client and server talking to each other are indeed talking to each other and not a third entity. Digital signature provides integrity and authenticity.

\par Web services and websites both work at the application layer. To use a website or web service a client first needs to connect with the server providing the service at the transport layer. This type of connection is called a TCP (Transmission Control Protocol) connection. TCP contains and carries the application layer data in the form of HTTP or SMTP inside it. There are quite a few Security mechanisms available at the transport layer that are quite robust and provide encryption, authentication and authorization. One such mechanism is Transport Layer Security (TLS) [2]. 

\par TLS makes use of two types of encryption algorithms: symmetric key encryption and asymmetric key encryption algorithms. In asymmetric key encryption, the server maintains two keys of which one is `public' and the other `private'. The server makes the public key available to all, and the private key is kept hidden. A message that is encrypted using the public key can only be decrypted using the private key. The client encrypts the message using this public key and sends it to the server. The server subsequently decrypts the message using the secret private key. In symmetric key encryption, both the server and client maintain a shared secret key between them called the symmetric session key. The symmetric session key is used both for encryption and decryption. To send a message, the client encrypts the message with this key and upon reaching the server, the same key is used for decryption. 

\par To transfer a large amount of data symmetric key algorithms are efficient as these are relatively light weight. To transfer fewer data asymmetric key algorithms are deemed more appropriate. The big question with symmetric key encryption is:  how to transfer the symmetric key initially from the client to server over a non-secure channel? A server provides its public key to all its clients. A client that wants to connect to the server provides the symmetric shared key information encrypted with the server's public key to the server. The server sees this shared key information after decrypting the client's message using the private key. After the server's agreement on  the shared key both start to transfer their data encrypted by the shared symmetric key. This is the normal sequence followed in encryption: first, the asymmetric key encryption, as described earlier, is used only to securely get the symmetric key for a session. Subsequent to this, in all further communications in that session between client and server symmetric key encryption is used.

\par Transport Layer Security (TLS), as described above, is normally used to encrypt the TCP container that contains the HTTP or SMTP container (application layer data) of the web service request and response. We know that both SOAP and RESTful web services contain XML, JSON etc inside the application layer packet (HTTP or SMTP), which is contained in the transport layer packet (TCP). TLS is effective with web services if the transfer of messages is between two parties only. In scenarios where several intermediate nodes need to access different parts of the content of the same HTTP or SMTP container, however, TLS becomes ineffective and the need arises for a message level security mechanism at the application layer [3]. This is a typical requirement of a web service composition scenario, where several nodes interact with each other to provide a larger composite application. 

\par For better comprehension of the scenario, let us consider a simple real world example of a big box containing a small box inside which there are several tennis balls with some relevant information written on them. Each ball belongs to a different person. The big box here can be looked upon as the HTTP container, the small box represents the entire XML or JSON message, and the balls represents various tags in XML or fields in JSON. The message written on one ball should be such that it cannot be understood by a person to whom it does not belong. What TLS  does is: it encrypts the  big box (Fig. 1.). When a person decrypts this big box, the small box and the information written on all the balls inside the small box become visible to that person irrespective of whether the ball belongs to him or not. Message level security, on the other hand, encrypts the information written on each ball and therefore only the person to whom a ball belongs is given the rights to decrypt the message. The balls here are the different tags of XML (or fields in JSON) that may be intended for different intermediate web service nodes in a web-service composition scenario. The entire XML should be encrypted so that the intermediate nodes are only able to understand and/or edit  tags that are meant for them. The need, therefore, is to encrypt the message written on each tag separately with different keys (Fig. 2.). Encrypting the various parts of an XML or JSON document with different keys is known as message level encryption. Message level encryption has the potential to provide message level security to web services. 

\begin{figure}[h!]   
\centering 
\includegraphics[width=1.5in,height=1.5in]{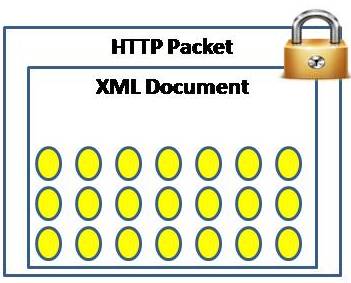}\\
\caption{\small \sl Encryption by TLS}
\label{fig1}
\end{figure}

\begin{figure}[h!]   
\centering 
\includegraphics[width=1.5in,height=1.5in]{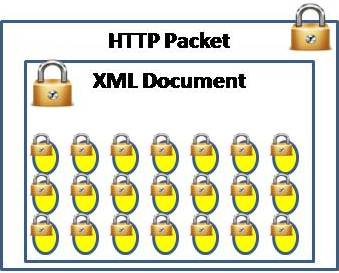}\\
\caption{\small \sl Message Level Encryption}
\label{fig1}
\end{figure}

\par SOAP based services do have formal mechanisms for message level security defined by W3C and OASIS [3] which provide authentication, authorization and encryption.  These services are broadly based on XML Signature [4] and XML Encryption                                                                      [5]. XML signature and XML encryption may work on the whole document, a part of the document, or even on binary documents. After encryption and signature, however, the encrypted document gets represented as XML only.

\par XML encryption uses existing block or stream ciphers to encrypt the document. Some of these algorithms are Advanced Encryption Algorithm (AES) [6], Data Encryption Algorithm [7] etc. Both XML encryption and signature use different versions of the Standard Hash Algorithm [8] for creating the message digest. For key agreement they use the Elliptic Curve Diffie-Hellman (ECDH) [9] Key Agreement. All these encryption and hash algorithms work on the data at the byte level. 

\par The issue with byte level encryption and signature is that two XML documents that represent the same data may not be identical byte-by-byte and as such even a slight cosmetic difference between the documents changes the whole encryption or signature. An example of this is: interchanging the position of attributes in a tag does not change the nature of the XML document, however, it completely changes the encryption and signature of that document. The normal procedure to overcome this problem is through XML cannonicalization  [10]. Here, several strict rules need to be applied on the  XML documents such that two XML documents that represent the same data become identical down to the last byte. Effective implementation of XML based signature and encryption requires the compulsory step of XML canonicalization.

\par This technique is ineffective if the two parties in conversation support different technologies, e.g. one supports XML content and the other supports JSON. Also, XML canonicalization is a compulsory step in XML based security and it is a cumbersome process indeed.  All this contributes to making such security techniques unsuitable for RESTful web services. Currently, there is no formal model for the provision of message level security in RESTful web services. In RESTful web services, both message encryption and authentic connection are still majorly dependent upon Transport Layer Security (TLS). 

\par Web service composition is a typical example where intermediate nodes need to access the same RESTful message and where TLS is not effective. In web service composition several service providers work together to form a large composite service offering. Each web service in a service composition does its bit and the larger composite task gets done. We are also moving towards a scenario where individual web services dynamically partake in a composition and after providing their respective services exit with the group. To realise such a scenario in a secure environment, it is imperative that an effective message level security mechanism be in place that would enable a web service only partial access to a message and keep the rest of it hidden.

\par We need to keep the various REST principles in mind while designing message level security models for RESTful services. In recent work by Gabriel Serme et. al [11] use of HTTP headers is made to transfer names of encryption algorithms, keys and other security meta data. Encryption and signature in this approach are done through the use of existing algorithms. We feel that sending new headers in HTTP packet makes the HTTP protocol itself non general. Further, use of existing encryption algorithms results in large sized HTTP contents.    

\par In this paper, we propose a novel approach to encrypting RESTful web services at the message level that works well simultaneously with  most popular content types (XML, JSON, HTML and plain-text). The main idea in this approach is to replace the ASCII printable characters with a series of numbers during encryption and doing the reverse during decryption. Encryption and decryption through this simple approach become easy and effective. This is because the encrypted message is quite small in size as compared to that of existing algorithms. In fact, in certain situations, the encrypted message is smaller than the message itself. The character to number conversion proposed also removes the need for XML cannonicalization because it is not a byte-by-byte encryption.  

\par The proposed approach also eliminates the need for content negotiation wherein the client and server negotiate on the type of content (XML, JSON) to be used for communication. This is because a resource gets converted into the same encryption irrespective of the type of representation. The series of numbers that constitutes the encryption will always be placed in the HTTP body as text/plain content type. The approach does not require special HTTP headers to pass security meta-data nor does it violate REST principles. 

\par Resource Representation is all about one to one mapping of XML tag-name and data type supported inside that tag, or data type supported by the value of the JSON name. Several web service providers represent their resources in their API documentation. In the proposed approach the server does not need to publish its resource representation. This is because all representations (XML, JSON etc.) get converted to the same encryption and can be decrypted to any representation. Data types in between the tags and JSON names are specified explicitly in the encryption itself. The client comes to know about resource representation in the encrypted form after its first request for the resource.  

\par It should be noted that in this paper we do not intend to provide a complete security solution for web service composition scenario. The idea is to introduce a novel, and much simpler  technique for encryption that is also secure, reliable, fast, and results in much smaller sized encrypted messages. The technique also removes the need for XML canonicalization and content negotiation.

\par The  paper  is  organized  as  follows:  in  Section 2,  we discuss the system architecture for our approach. In Section 3 we describe our proposed approach in detail with a running example for better understanding. We discuss various advantages of our approach in Section 4 and conclude the paper in Section 5.

\section{System Architecture}
\label{S:2}
\par In RESTful communications and in general XML-tags, XML-attributes, values of XML-attributes and JSON-names change less frequently and therefore we call these the \textit{non-variable} parts of the request and response messages. The values in between XML-tags or the values of JSON names on the other hand change much more frequently and we call these the \textit{variable} parts of request and response messages. For example:

\textbf{XML 1:}
\begin{verbatim}
<root attr1="value1" attr2="value2">
<name>iiti</name>
<value>2</value>
</root>
\end{verbatim}

\textbf{JSON 1:}
\begin{verbatim}
"root": {
    "-attr1": "value1",
    "-attr2": "value2",
    "name": "iiti",
    "value": "2"
}
\end{verbatim}

XML and JSON can be easily converted into each other. An XML and its corresponding JSON are shown above as XML 1 and JSON 1. In XML 1 the ``root", ``name", ``value", ``attr1", ``attr2", ``value1" and ``value2" are the \textit{non-variable} parts and ``iiti" and ``2" are the \textit{variable} parts of the message. In the corresponding JSON message JSON 1, we have similar \textit{non-variable} and \textit{variable} parts. 

\par In our approach of encryption both the \textit{non-variable} and \textit{variable} parts of the message are represented by a string of numbers. The encrypted message is sent to the receiver encapsulated within an HTTP packet as plain-text. We know that XML and JSON can be easily converted to each other and as stated earlier the strength of  this algorithm is that it does not matter whether the content to be encrypted is in JSON or XML the result is the same encrypted message which can subsequently be decrypted into JSON and/or XML as the need be. Encryption and decryption can be done by a shared symmetric key among parties. In case of web service composition where different intermediaries have access to the same RESTful message, different parts of the same message can be encrypted with different symmetric keys. We initially discuss message level encryption between two parties only. Subsequently, in Section 3 we escalate the discussion to multiple party web-service composition.  

\par Figure 3 shows the various components at the client and server ends that are used for encryption/decryption.  At the server side,  there is a message level encryption/decryption module (Encryption-Decryption Engine) that lies between the RESTful service application and the HTTP service. At the client side, the same lies between the RESTful client and the HTTP client. There is a `Key Manager' at both ends that is connected to the corresponding HTTP server and client.   


\begin{figure}[h!]   
\centering 
\includegraphics[width=3.5in,height=3.0in]{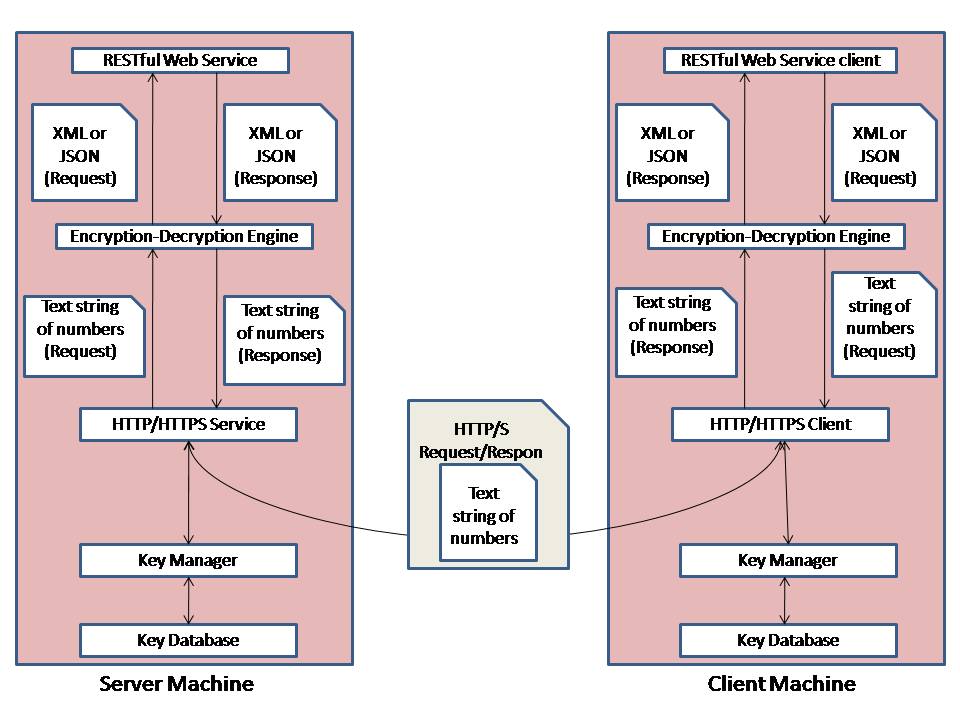}\\
\caption{\small \sl Cliet Server Interaction}
\label{fig1}
\end{figure}


\subsection{Key Manager}
\par Prior to the start of a conversation, both ends require a symmetric session key. The key manager at both ends manages the symmetric key. First, the client's key manager sends a key request to the server's key manager as a special command (``Get key"). The ``Get key" command goes to the HTTP client section at the client side. The HTTP client section at the client side places the ``Get key" command in the body of the HTTP POST request with the content type being plain-text as shown in Figure 4.

\begin{figure}[h!]   
\centering 
\includegraphics[width=3.5in,height=1.5in]{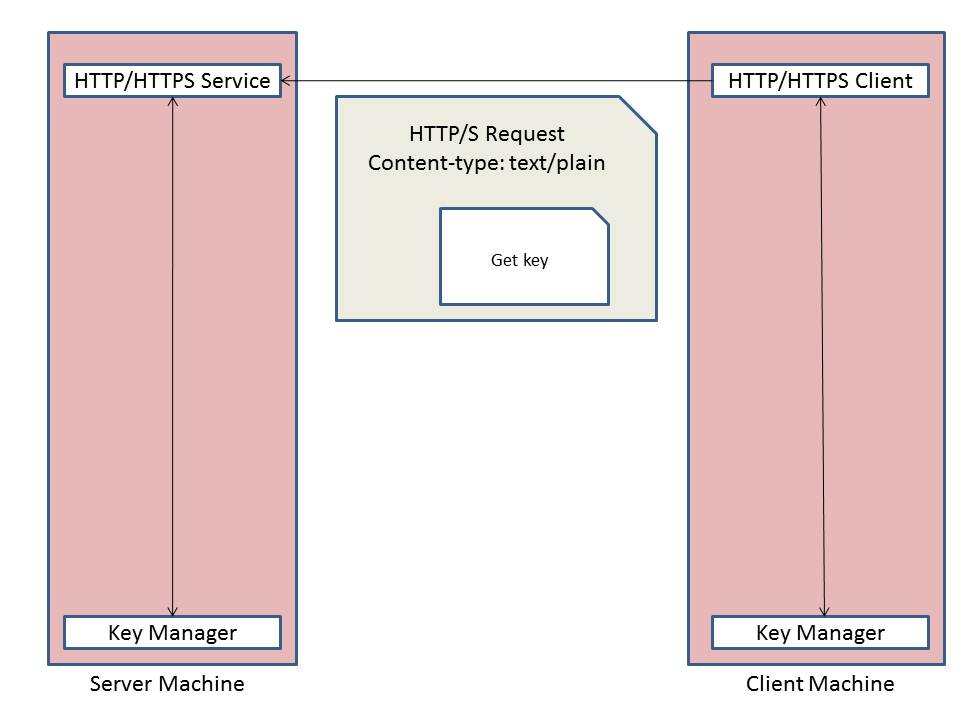}\\
\caption{\small \sl Client's request for key}
\label{fig1}
\end{figure}

The HTTP service at the server side receives the request and finds the ``Get key" string in the POST request's body. Subsequently, the server's key manager generates 10 random integers within a given range and creates a unique \textit{10 element key}. The server's key manager passes this \textit{10 element key} to the HTTP service at the server side. The HTTP service at the server side creates an HTTP response, places the key in the response body with the content type being plain text and sends the HTTP response to the HTTP client at the client side (Fig. 5). The HTTP client at the client side reads the response body and passes the content to the client's key manager.  

\begin{figure}[h!]   
\centering 
\includegraphics[width=3.5in,height=1.5in]{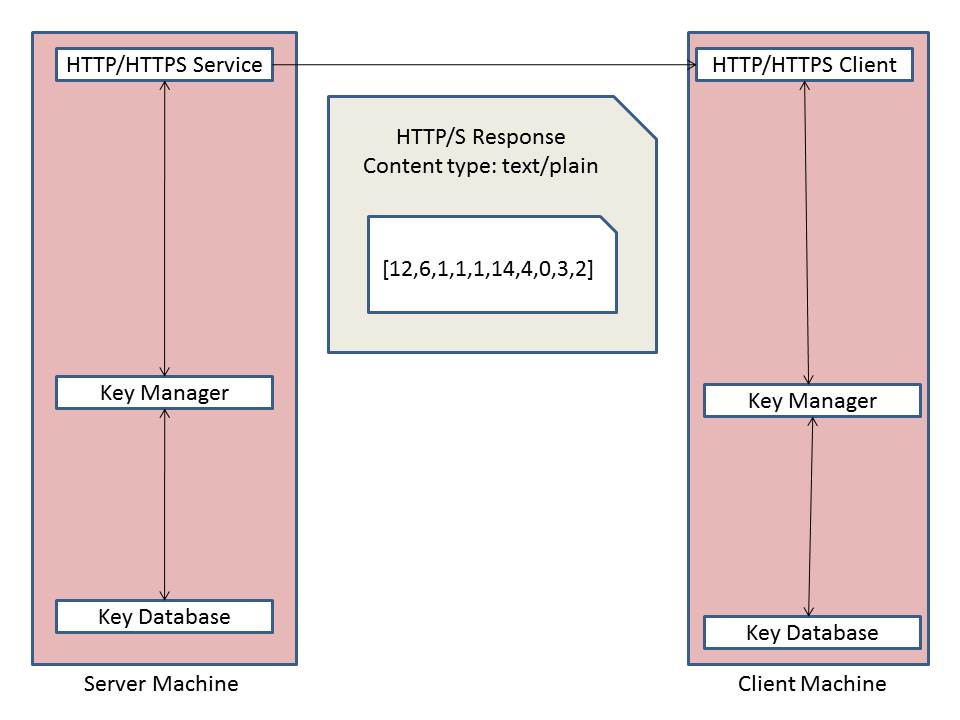}\\
\caption{\small \sl Server's response with key}
\label{fig1}
\end{figure}

The conversation between the client and the server for a key is a matter between just two parties and therefore the key exchange can utilise TLS for security. Both sides store the shared key in a key database for further use. The number of keys maintained by each member in a web services composition group depends on the structure of the web services composition. At most, each member in a group of n service providers may need to maintain n secret keys (n-1 keys to communicate with the other n-1 members and one group key). The shared key gets stored in key-databases of the client and the server. The client is responsible for a change in key. To change the key the client sends a new key request to the server.

\subsection{10 Element Key}    
\par  This symmetric key is at the heart of the proposed approach and comprises 10 elements. Using just this key both the client and server can create various tables that take part in message level encryption and decryption. We have seen an example of the \textit{10 element key} randomly generated by the server in Figure 5 \textit{([12,6,1,1,1,14,4,0,3,2])}. In general, the key is represented as \textit{key} = \textit{[rows, cols, start\_with, row\_rev, col\_rev, symbol\_type, group\_size, reverse, final\_sum, power]}. In this paper different elements of the \textit{key} are referred to by using indexing starting from \textit{0}, for example \textit{key[0]} for \textit{rows}, \textit{key[1]} for \textit{cols} etc. We will now understand the various elements of the key one by one.

\subsubsection{rows (key[0])} The Encryption-Decryption engine at both the client and server sides maintains a table that we choose to call the \textit{Temporary Table (TT)}. \textit{rows} at \textit{key[0]} specifies the number of rows in this table. The number of rows in \textit{TT} = \textit{key[0]} + \textit{1}. \textit{key[0]} ranges from \textit{1} to any integer depending upon the capability of the system.

\subsubsection{cols (key[1])} \textit{cols} at \textit{key[1]} specifies the number of columns in the \textit{TT}.  The number of columns in \textit{TT} = \textit{key[1]} + \textit{1}. For \textit{key = [12,6,1,1,1,14,4,0,3,2]} the \textit{TT} at both the client and server sides will have \textit{13} rows and \textit{7} columns (including the coloured `header' rows and columns) as shown in Figure 6. The  coloured cells of the table are used to store table headers, whereas the white cells are for normal entries. \textit{key[1]} ranges from \textit{1} to any integer depending upon the capability of the system. 

\begin{figure}[h!]   
\centering 
\includegraphics[width=1.5in,height=1.5in]{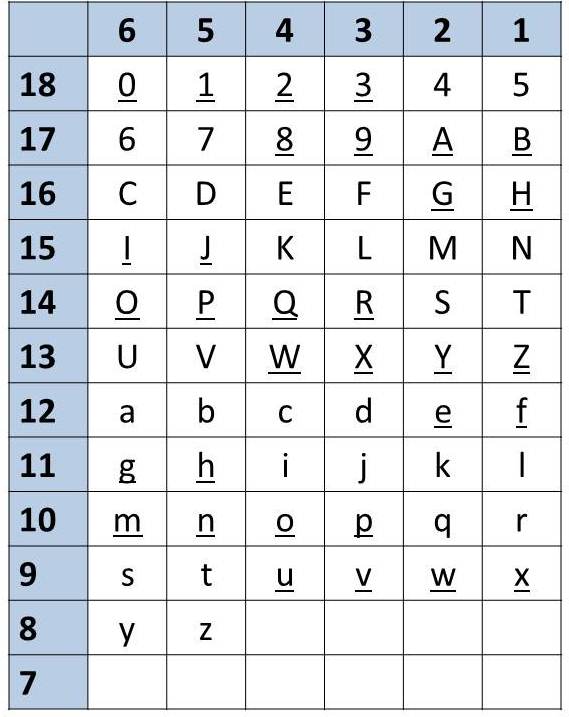}\\
\caption{\small \sl Temporary Table (TT)}
\label{fig1}
\end{figure}

\subsubsection{start\_with (key[2])} We have two types of headers in \textit{TT}, one is a \textit{row header (rh)} and the other is the \textit{column header (ch)}. As part of the encryption process, we need to number both \textit{rh} and \textit{ch}. Numbering always starts with \textit{1}. Numbering can start from either \textit{rh} or \textit{ch} depending upon the value of \textit{key[2]}. The header numbering of \textit{TT} starts with \textit{rh} if the value of \textit{key[2]} is \textit{0} and it starts with \textit{ch} if the value of \textit{key[2]} is \textit{1}. Range of \textit{key[2]} is \{0, 1\}. If the numbering of the header starts with numbering \textit{rh}, then the same count continues while numbering  \textit{ch} and vice-versa. For example, in \textit{key = [12,6,1,1,1,14,4,0,3,2]} the header numbering starts with \textit{ch}. In Figure 6, therefore, the count \textit{1} to \textit{6} has been used for numbering \textit{ch} and in continuation with that \textit{7} to \textit{18} have been used for numbering \textit{rh}.

\subsubsection{row\_rev (key[3])} If \textit{key[3]} is \textit{0} then the \textit{rh} numbering starts from the top most row (top to bottom) and if \textit{key[3]} is \textit{1} then the \textit{rh} numbering starts from the bottom row (bottom to top). For \textit{key = [12,6,1,1,1,14,4,0,3,2]} the \textit{rh} numbering starts from the bottom and progresses towards the top as shown in Figure 6. The range of \textit{key[3]} is \{0, 1\}.

\subsubsection{col\_rev (key[4])} If \textit{key[4]} is \textit{0} then the \textit{ch} numbering starts from the left most column (left to right) and if \textit{key[4]} is \textit{1} then the \textit{ch} numbering starts from the right most column (right to left). For \textit{key = [12,6,1,1,1,14,4,0,3,2]}, the \textit{ch} numbering starts from the right and progresses towards the left as shown in Figure 6. The range of \textit{key[4]} is \{0, 1\}.

\subsubsection{symbol\_type (key[5])} There are four types of characters that can be used in this table as elements in the non-header cells: small alphabet, capital alphabet, digits, and special symbols. The total possible arrangements in this table are: $^4C_1\times{(1!)} + ^4C_2\times(2!) + \\^4C_3\times(3!) + ^4C_4\times(4!)$ = 64. For all these 64 arrangements \textit{key[5]} ranges from 0 to 63. Each value defines a unique occurrence, non occurrence, and order of occurrence of the various printable ASCII characters in the non header cells of the \textit{TT}. For each value of \textit{key[5]}, the entry of characters in \textit{TT} starts from the top left corner non-header cell and sequentially moves towards the bottom right corner non-header cell. For example, 1) If value \textit{key[5]} is \textit{0} then only small alphabet characters are allowed in \textit{TT}, 2) If it is \textit{1} then only capital alphabet characters are allowed in \textit{TT}. Similarly, if \textit{key[5]} is \textit{14} then the order of entering characters in \textit{TT} is \textit{digits}, followed by \textit{capital alphabets}, further followed by \textit{small alphabets}. The value \textit{14} ignores special symbols. There are various other arrangements of characters for other values of \textit{key[5]}. Another value of \textit{key[5]} may have included special symbols and the various characters may have been in a different order of occurrence. Figure 6 shows the entry of \textit{TT} if \textit{key = [12,6,1,1,1,14,4,0,3,2]}. \textit{key[5]} and the corresponding arrangement of characters are only assumptions of the authors. During implementation, these may be different, but the total possible combination is always fixed.

\subsubsection{group\_size (key[6])} Various groups of consecutive elements in \textit{TT} are created. Each group has a size indicated by \textit{key[6]}. The last group of \textit{TT} can have fewer elements than \textit{key[6]}. For \textit{key = [12,6,1,1,1,14,4,0,3,2]} since \textit{key[6]} is \textit{4}, \textit{TT} has various groups of \textit{4} elements each. In Figure 6 we represent the various groups as alternately underlined and non-underlined elements. The elements \textit{0,1,2,3} are in the same group and \textit{4,5,6,7} are in a different group, similarly elements \textit{8, 9, A, B} are in the same group and \textit{C, D, E, F} are in a different group, and so on. \textit{key[6]} ranges from \textit{1} to \textit{(key[0]-1) $\times$ (key[1]-1)}.

\subsubsection{reverse (key[7])} If \textit{key[7]}  is 1 then the elements in all the groups will be in a group wise reverse order with respect to their initial positions in the group. If this value is 0 then no such reverse operation occurs. For \textit{key = [12,6,1,1,1,14,4,0,3,2]}, the \textit{TT} is shown in Figure 6. For \textit{key = [12,6,1,1,1,14,4,1,3,2]} the groups get reversed and the corresponding \textit{TT} is shown in Figure 7.  

\begin{figure}[h!]   
\centering 
\includegraphics[width=1.5in,height=1.5in]{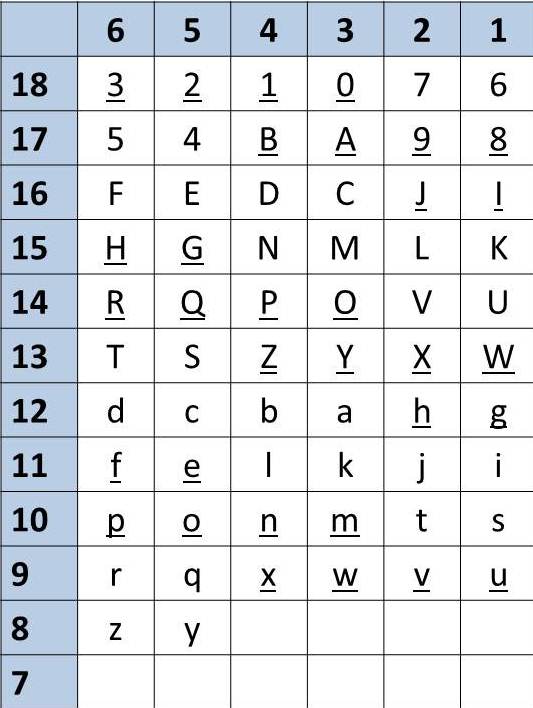}\\
\caption{\small \sl Temporary Table (TT)}
\label{fig1}
\end{figure}

\subsubsection{final\_sum (key[8])} \textit{key[8]} will be explained later. 

\subsubsection{power (key[9])} In the process of message level encryption,  we need to calculate a unique integer corresponding to each character present in \textit{TT}. This integer may be calculated as \textit{$(rh)^{key[9]}+(ch)^{key[9]}$}. For example, for the following \textit{key = [12,6,1,1,1,14,4,1,3,2]} the integer value of \textit{G} is \textit{$(15)^{2}+(5)^{2} = 250$}, similarly the integer value for \textit{9} is \textit{$(17)^{2}+(2)^{2} = 293$}, and so on. It is possible that two different characters may end up in the same integer, this situation is called integer collision. If integer collision occurs then we keep on increasing the integer value of the later characters in \textit{TT} until it gets a unique value. Powering \textit{ch} and \textit{rh} with \textit{key[9]} has two benefits. 1). It makes the encryption itself more random and 2). The Probability of integer collision becomes very less. Range of \textit{key[9]} from \textit{1} to any integer depends on the capabilities of the service provider and consumer.

\par It is important to note that the rules outlined for the Tag Table are for demonstrating the idea and are open to modification. The main idea is to bring in as much randomness as possible so as to make it impossible for an adversary to guess the contents.

\subsection{Encryption and Decryption engines}

\par A client's PUT or POST request that may be in XML, JSON or plain text is first sent to the Encryption-Decryption engine. In the Encryption-Decryption engine at the client side, the request message gets converted into a string of numbers. Subsequent to this, the HTTP/S client encapsulates this string of numbers within an HTTP/S request (PUT or POST) packet and sends it to the server. At the server end, the HTTP/S server reads the string of numbers and sends it to the Encryption-Decryption engine at the server end. The Encryption-Decryption engine at this end decrypts the string of numbers into the corresponding XML and/or JSON form and finally delivers it to the Web service. The same process repeats with the response from the server to the client. The Encryption and decryption engines used here comprise the following sub-components:

\subsubsection{Temporary Table (TT)} Temporary Tables (TT) have been discussed at length in the earlier sub-sections. We assume that the contents of this table comprise only printable ASCII characters. Each element in the table, comprising printable ASCII characters, is uniquely defined by a corresponding pair of \textit{rh} and \textit{ch}. The Structure of the table and the position of various characters in the cells is established by the parties in conversation using the symmetric key as discussed earlier. The \textit{TT} is deleted after the creation of the symbol table.
  
\subsubsection{Symbol Table (ST)} This comprises a table with two columns without a row or column header. The table maps each printable ASCII character to a unique integer derived from \textit{TT} using the symmetric key. The table is mainly used for special encryption and decryption of ASCII printable RESTful content called \textit{symbol table based encryption (STBE)} and \textit{symbol table based decryption (STBD)}. The Creation of \textit{ST}, its purpose, and the various steps of \textit{STBE} and \textit{STBD} are covered in the next section.

\subsubsection{Tag Table (TAT)} This component includes a table with two columns without a row or column header. The table maps the \textit{non-variable} parts of a RESTful message (XML, JSON or HTML) to a unique integer. The table is used for another encryption and decryption on ASCII printable RESTful contents called \textit{tag table based encryption (TATBE)} and \textit{tag table based decryption (TATBD)}. The creation of \textit{TAT}, its purpose and the various steps involved in \textit{TATBE} and \textit{TATBD} are covered in the next section. 

\section{The Approach}
\label{S:3}
\par Before further discussion we assume that the \textit{10 element key} (\textit{key = [12,6,1,1,1,14,4,1,3,2]}) has already been exchanged between the two parties and \textit{TT} has been created at both ends. In the following subsections, the remaining encryption procedure is discussed  based on the \textit{10-element key} and the \textit{TT} created based on this key (Fig. 7). We will be using a running example for better understanding of the approach. Figure 8 summarises our approach. The figure describes both encryption and decryption.

\begin{figure}[h!]   
\centering 
\includegraphics[width=3.0in,height=3.0in]{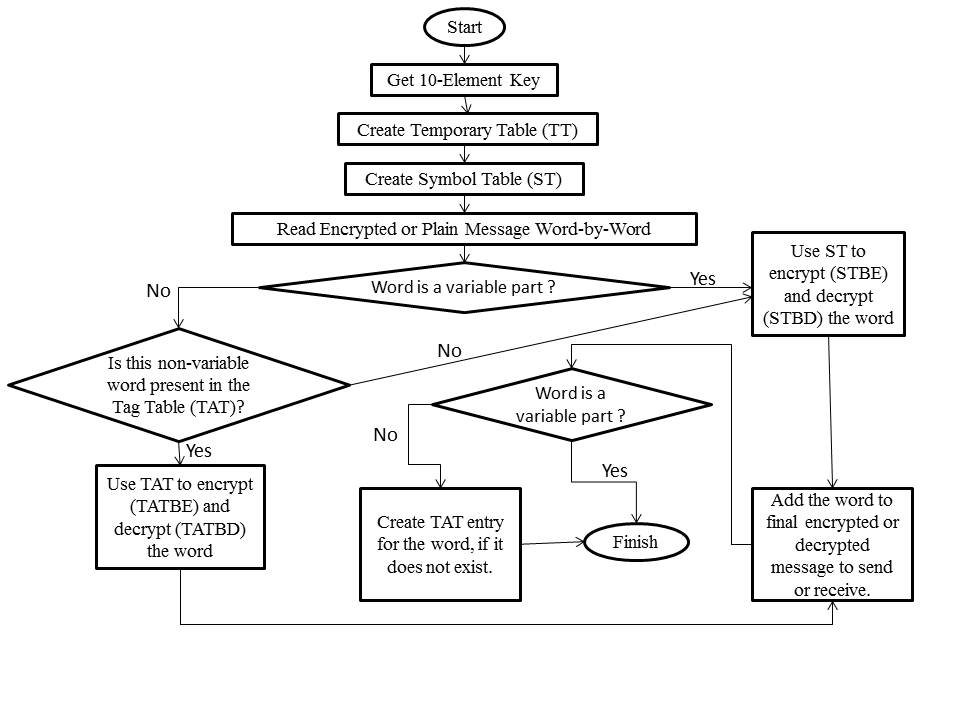}\\
\caption{\small \sl Flow Chart of our Approach}
\label{fig1}
\end{figure}

\par Our purpose is to provide a substitution cipher which converts the plain-text into a series of numbers. Such substitution cipher removes the need of XML canonicalization. As shown in Figure 8, after the creation of \textit{TT}, the Symbol Table (\textit{ST}) gets created as shown in Figure 9. Refer to the various \textit{variable} and \textit{non-variable} parts of both encrypted and plain messages as
words. The symbol table maps each character in a non-header cell of the \textit{TT} to a unique integer. We use this table to replace
each character in the XML or JSON documents to a corresponding unique integer. For example, the word \textit{``iiti"} gets converted into
\textit{``122122104122"}. This is called \textit{ST} Based Encryption i.e \textit{STBE}. Decrypting \textit{``122122104122"} back into \textit{``iiti"} is called \textit{ST} based decryption (\textit{STBD}). The process of \textit{ST} creation, \textit{STBE} and \textit{STBD} are discussed in the following sub-sections. Since the \textit{ST} can be created based on the \textit{10-element} key, both parties can start encryption and decryption using \textit{ST} without knowing the structure of the message. Using \textit{STBE} results in a large size encrypted message because each character either in the variable or non-variable parts gets converted to a unique integer. This is why we introduced the concept of Tag Table (\textit{TAT}). The \textit{TAT} maps the whole \textit{variable} parts of the message to a unique integer (Fig. 10). Using the Tag Table Based Encryption (\textit{TATBE}) results in an encrypted message size that is smaller than the \textit{STBE}. For example, the tag \textit{``root"} can be converted into \textit{``04"} using the \textit{TATBE} instead of \textit{``0117126126104"} using the \textit{STBE}. Decrypting \textit{``04"} back in \textit{``root"} is called the Tag Table Based Decryption (\textit{TATBD}). The process of the \textit{TAT} creation, \textit{TATBE} and \textit{TATBD} are discussed in the following subsections. The idea is to use the \textit{STBE} and \textit{STBD} for \textit{variable} parts and for those \textit{non-variable} parts of the message whose \textit{TAT} entries still do not exists on either side. When the \textit{STBE} and \textit{STBD} are being used for the \textit{non-variable} parts, the Tag Table (\textit{TAT}) entry for the same must be created for later use. Use \textit{TATBE} for those \textit{variable} parts whose entries are present in the \textit{TAT} at both sides.

\subsection{Symbol Table Creation}

\par The Symbol Table (\textit{ST}) gets created by the client and server based on the Temporary Table (\textit{TT}) in the following manner: The \textit{ch} and \textit{rh} of all \textit{non-header} characters inside the \textit{TT} are identified and the corresponding unique number for each \textit{non-header} symbol in the same is calculated. The procedure for creating \textit{ST} is shown below:

\begin{enumerate}
  \item For each \textit{non-header} and \textit{non-empty} cell in \textit{TT}, the printable character is taken out and its corresponding unique integer value is calculated as \textit{value} = $(rh)^{key[9]}$+$(ch)^{key[9]}$. 
  \\For example \textit{TT} in Fig. 7,  character \textit{`j'} has \textit{rh} = \textit{11} and \textit{ch} = \textit{2}. The calculated \textit{value = $(11)^{2}$+$(2)^{2}$}. So \textit{value} corresponding to \textit{`j'} in \textit{TT} is \textit{125}.
  
  \item The number of digits in \textit{value} is made the same as the value of \textit{key[8]}. In doing this, we first calculate \textit{diff} = \textit{key[8]}-\textit{(no. of digits in value)} and then proceed to step 3. 
   \\For example if \textit{key = [12,6,1,1,1,14,4,1,3,2]} (\textit{key[8]} is \textit{3}) then \textit{diff} = \textit{3}-\textit{3} = \textit{0}.
 
  \item We calculate the final value that is to be inserted in \textit{ST} corresponding to the given \textit{non-header} character of \textit{TT}. We do it as \textit{final\_value} = \textit{value}$\times10^{diff}$. 
  \\For example \textit{final\_value} of \textit{`j'} is \textit{125$\times10^{0}$} = \textit{125} .

  \item The final record to be inserted in \textit{ST} is the \textit{(character, final\_value)} pair. The \textit{final\_value} is the primary key of \textit{ST}. If a record in \textit{ST} exists with the same \textit{final\_value} then go to step 5 otherwise, go to step 6 .
   \\For example \textit{(`j',125)} is going to be inserted in \textit{ST}. Before insertion of this record, we first check if a record in \textit{ST} exists with the same \textit{final\_value} (125). For now, we assume that no such record exists, and therefore  go to step 6. If \textit{(`j',125)} is already in \textit{ST} and  \textit{(`o',125)} has to be inserted we go to step 5.
   
   \item When a record with the same \textit{final\_value} already exists in \textit{ST}, the \textit{final\_value} of the record is incremented by \textit{1} and checked for uniqueness. This is continued until a unique \textit{final\_value} is found. Subsequently, step 6 is executed. Note that even if we considered  \textit{final\_value} as a primary key more than one entry of a \textit{character} is not possible, because each character is unique in \textit{TT}.
   \\For example before inserting a record for character \textit{`o'}, we increase its \textit{final\_value} to \textit{126}. If a record with \textit{final\_value} \textit{126} also exists then we make it \textit{(`o',127)}, otherwise, we record \textit{(`o',126)} and proceed  to step 6.
   
   \item Insert the record \textit{(character, final\_value)} in \textit{ST}.     
\end{enumerate}

  \par A subset of \textit{ST} is shown in Fig.9.
 
\begin{figure}[h!]   
\centering 
\includegraphics[width=1.5in,height=2.0in]{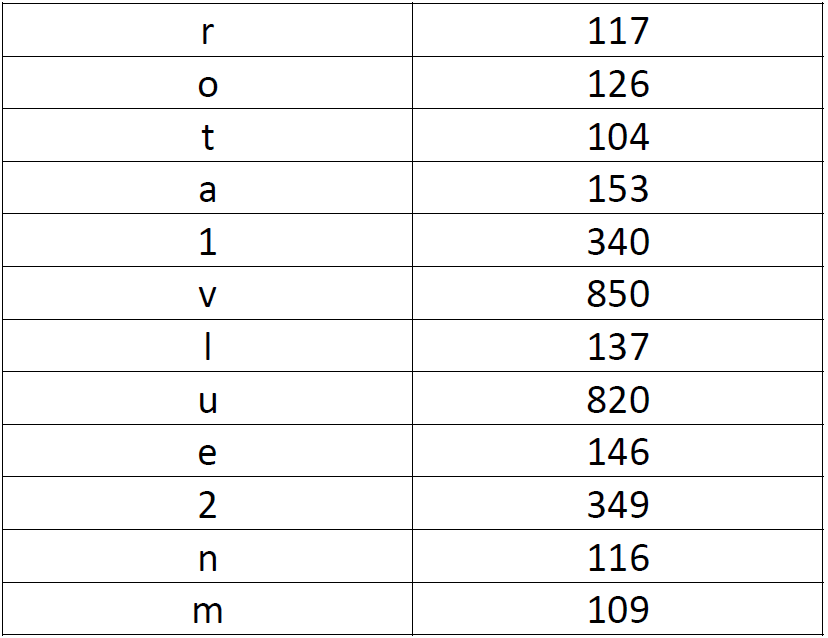}\\
\caption{\small \sl A Subset of ST}
\label{fig3}
\end{figure}

\subsection{Symbol Table Based Encryption (STBE) and TAT Creation at Server Side}
By now both the client and server should have the same Temporary Table (\textit{TT}) and Symbol Table (\textit{ST}). The client now sends its first request to the server. The requested resource would comprise several \textit{variable} and \textit{non-variable} parts (either XML or JSON). The \textit{non-variable} part could be any name and \textit{variable} part could be of any data-type.  Until this point, the client only knows the URI of the resource and does not know anything about the resource representation. If the client makes an HTTP request that does not require an HTTP body (e.g GET) then the server responds after encrypting the response message using \textit{ST} as \textit{STBE}. The client decodes this using the Symbol Table Based Decryption (\textit{STBD}). The decrypted message gets converted into XML or JSON. In the decrypted document the name of the \textit{non-variable} parts and data type of the \textit{variable} parts will be clear to the client. We will explore \textit{STBD} in a later sub-section. 

\par If the client request requires an HTTP body (e.g PUT, POST) and the client does not have any idea about the resource representation in the request body (i.e what to send in the request body) then the client will send an empty POST request on the URI. In such a case, the server sends an encrypted resource representation using the \textit{STBE}. The \textit{STBE} is used by the server when the client is not in the know of the \textit{non-variable} parts of the message. As soon as the client becomes aware of the \textit{non-variable} parts, the server stops encrypting the message using the \textit{STBE} and starts encrypting it using the Tag Table Based Encryption (\textit{TATBE}). We will explore \textit{TATBD} in a later sub-section.

\par In the eventuality that the server introduces a new \textit{non-variable} part in between a conversation, the new \textit{non-variable} part is encrypted using \textit{STBE} and the encryption of the rest of the \textit{non-variable} parts is done through \textit{TATBE}. The \textit{variable} parts of the message always get encrypted using \textit{STBE}. The client never needs to encrypt the \textit{non-variable} parts of its  request body using the \textit{STBE}. This is because the client can never introduce a new \textit{non-variable} part in the request. The client, however, needs to decrypt the \textit{non-variable} parts of the response sent out by the server that is encrypted using \textit{STBE}. The process of encrypting \textbf{XML1} using \textit{STBE} and creation of Tag Table (\textit{TAT}) for each \textit{non-variable} parts of the message happens simultaneously at the server side is depicted in the steps below.  
    
\par Lets assume that the encrypted message gets stored in a string variable \textit{STEnc}. The initial value of \textit{STEnc} is \textit{null}. We have two more variables \textit{noOfNonVars} and \textit{noOfDigitsForNonVars}, both are initially \textit{0}. The \textit{variable} and \textit{non-variable} parts of the XML are taken in the same order that they appear in the document. We together call these parts (the \textit{variable} and \textit{non-variable} parts of the XML) a \textit{word}. For example in \textbf{XML1} the order of occurrence of the words are \textit{``root"}, \textit{``attr1"}, \textit{``value1"}, \textit{``attr2"}, \textit{``value2"}, \textit{``name"}, \textit{``iiti"}, \textit{``/name"}, \textit{``value"}, \textit{``2"}, \textit{``/value"} and \textit{``/root"}.

\par In the case of web service composition different tags may belong to different service providers. Different tags are therefore encrypted using different keys. The receiver must be notified about tag numbers which have been encrypted by its key. There are therefore comma separated tag numbers appended at the beginning of \textit{STEnc} followed by space. For a tag number present at the beginning of \textit{STEnc}, the client has access to all child elements of the given tag.
\\For example \textit{(1,)} at the beginning of \textit{STEnc} says that the receiver has access to the tag \textit{``root"} and its children in \textbf{XML1} (i.e whole document). If \textit{(2,)} present at the beginning of \textit{STEnc} says that the receiver has access to the tag \textit{``name"} and its children in \textbf{XML1}. The presence of \textit{(2,3,)} at the beginning of \textit{STEnc} says that the receiver has access to the tags \textit{``name"} and \textit{``value"} and their children in \textbf{XML1}. Normally common tags like \textit{``root"} get encrypted using an agreed upon group-key among members. 
	
	\par If word \textit{(1,)} gets added in STEnc, then STEnc which was initially \textit{null} becomes: 
	\\ \textit{STEnc = ``1, "}. 
\begin{enumerate}
	\item Repeat step-2 to step-7 for each word in the XML. 
	\item If the word is a tag then the same is converted into a string of numbers preceded by \textit{0} using the \textit{ST} reference for each character of the word. This string of numbers is then concatenated into a single string \textit{STEnc} followed by a space and step 7 is executed.
	\\For example, the first \textit{word} of \textbf{XML1} is a tag \textit{``root"}. This consists of four characters \textit{`r'}, \textit{`o'}, \textit{`o'} and \textit{`t'}. The \textit{ST} entries corresponding to these characters are \textit{`117'}, \textit{`126'}, \textit{`126'} and \textit{`104'} respectively. These numbers get concatenated as  \textit{``0117126126104"}. Therefore, the encrypted form of the tag \textit{root} along \textit{STBE} is \textit{``0117126126104"}. Concatenate the  encrypted form of the word in \textit{STEnc} followed by a space. If STEnc was initially \textit{``1, "} and then it gets updated as:
\\ \textit{STEnc = ``1, 0117126126104 "}.

	\item If the word is an attribute-name, the same is converted into a string of numbers preceded by \textit{00} using the \textit{ST} reference for each character of the word. This string of numbers is then concatenated into \textit{STEnc} followed by a space and step 7 is executed.
	\\For example, the second \textit{word} of \textbf{XML1} \textit{``attr1"} is an attribute-name. This consists of five characters \textit{`a'}, \textit{`t'}, \textit{`t'}, \textit{`r'} and \textit{`1'}. The Symbol Table \textit{ST} entries corresponding to these characters are \textit{`153'}, \textit{`104'}, \textit{`104'}, \textit{`117'} and \textit{`340'} respectively. These numbers are concatenated as  \textit{``00153104104117340"}. The encrypted form of the attribute-name \textit{attr1} along \textit{STBE} is therefore \textit{``00153104104117340"}. Concatenate the encrypted form of the word in \textit{STEnc} followed by a space. If STEnc was initially \textit{``1, 0117126126104 "}, it gets updated as:
\\ \textit{STEnc = ``1, 0117126126104 00153104104117340 "}.

	\item If the word is an attribute-value, the same is converted into a string of numbers preceded by \textit{000} using the \textit{ST} reference for each character of the word. This string of numbers is then concatenated into \textit{STEnc} followed by a space and step 7 is executed.
		\\For example, the third \textit{word} of \textbf{XML1} \textit{`value1"} is an attribute-value. This consists of six characters \textit{`v'}, \textit{`a'}, \textit{`l'}, \textit{`u'}, \textit{`e'} and \textit{`1'}. The \textit{ST} entries corresponding to these characters are \textit{`850'}, \textit{`153'}, \textit{`137'}, \textit{`820'}, \textit{`146'} and \textit{`340'} respectively. These numbers get concatenated as \\ \textit{``000850153137820146340"}. Therefore, the encrypted form of the attribute-value \textit{value1} along \textit{STBE} becomes \textit{``000850153137820146340"}. As earlier, this is concatenated with \textit{STEnc} followed by a space. If STEnc was initially \textit{``1, 0117126126104 00153104104117340 "}, it gets updated as:
\\ \textit{STEnc = ``1, 0117126126104 00153104104117340 000850153137820146340 "}.
		
		\item \textit{0} is used to represent a closing tag. Whenever a closing tag is encountered \textit{0} is concatenated with the string \textit{STEnc} followed by a space.
		\item If the word is a \textit{variable} part of the XML the same is converted into a string of numbers using the \textit{ST} reference for each character of the word. This string of numbers is then concatenated into \textit{STEnc} followed by a space. 
		\\For example, the seventh \textit{word} of \textbf{XML1} \textit{``iiti"} is a \textit{variable} part of the XML . This consists of four characters \textit{`i'}, \textit{`i'}, \textit{`t'} and \textit{`i'}. The \textit{ST} entries corresponding to these characters are \textit{`122'}, \textit{`122'}, \textit{`104'} and \textit{`122'} respectively. These numbers are concatenated as  \textit{``122122104122"}. This string of numbers is then concatenated into \textit{STEnc} followed by a space.
		After completion of the encryption of the whole \textbf{XML 1}, the encrypted message \textit{STEnc} looks like this:
\\ \textit{STEnc} = \textit{``1, 0117126126104 00153104104117340 000850153137820146340 00153104104117349 000850153137820146349 0116153109146 122122104122 1 0 0850153137820146 349 2 0 0"}.
	  \\Since \textit{key[8]} decides the size of the Symbol Table Based Encrypted message, the range of \textit{key[8]} is from \textit{1} to an integer depending on the network latency between the service provider and the client.
		\item This step is the \textit{TAT} creation step. This step is arrived at only for \textit{non-variable} parts of the XML. The numbers corresponding to each character in the \textit{non-variable} words are added and stored in the variable \textit{sum}.  The number of digits in the integers representing the \textit{non-variable} parts of the XML (say \textit{noOfDigitsForNonVars}), should be minimum and such that it can accommodate all \textit{non-variable} parts in it. It can be decided by the server and client separately and automatically based on the number of \textit{non-variable} parts currently in use for communication (say \textit{noOfNonVars}). The following calculations are next done:    

\begin{enumerate}
\item \textit{noOfNonVars} can be calculated as a summation of the number of new   \textit{non-variable} parts introduced in the current message and the number of entries already present in the \textit{TAT}. 
\item \textit{noOfDigitsForNonVars}= $\ceil{\log _{10} \left( noOfNonVars\!+1 \right)}$
\par Here we add \textit{1}, because \textit{noOfNonVars} as a multiple of \textit{10} gives a wrong impression about the \textit{noOfDigitsForNonVars}, if \textit{noOfDigitsForNonVars} = $\ceil{\log _{10} \left( noOfNonVars \right)}$. This is because we are using \textit{zero} for indicating the type of the word, therefore we should not include \textit{zero} in the \textit{non-variable} part of the number mapping. If \textit{noOfNonVars} is \textit{10} then $\ceil{\log _{10} \left( noOfNonVars \right)}$ results in \textit{1}. This implies that in such a case we need to necessarily use all digits from \textit{0} to \textit{9} to map the various \textit{non-variable} parts in \textit{TAT}. Adding \textit{1} in the above formula removes this limitation.    	
\item \textit{diff} = \textit{noOfDigitsForNonVars}-\textit{(no. of digits in sum)}
\item \textit{sum} = \textit{sum}$\times$\textit{$10^{diff}$} 		
\item Finally, the record \textit{(word, sum)} is stored in \textit{TAT}. However if the corresponding \textit{sum} is already present in \textit{TAT} then repeat:  \textit{sum} = (\textit{sum}+\textit{1})\% \textit{$10^{noOfDigitForNonVars}$} until \textit{sum} gets a unique value other than \textit{0}.
\end{enumerate}

\par For example, the tag \textit{``root"} consists of four characters \textit{`r'}, \textit{`o'}, \textit{`o'}, \textit{`t'}. The \textit{ST} entries corresponding to these characters are \textit{`117'}, \textit{`126'}, \textit{`126'} and \textit{`104'} respectively. The steps followed for the word \textit{``root"} are:

\begin{enumerate}
\item \textit{sum} =  \textit{117} + \textit{126} + \textit{126} + \textit{104} = \textit{473}. 
\item \textit{noOfNonVars} = number of new \textit{non-variable} parts introduced in the current message (\textit{7}) + number of entries already present in the \textit{TAT} (\textit{0}) = \textit{7}.
\item  \textit{noOfDigitsForNonVars} = $\ceil{\log _{10} \left( 7+1 \right)}$ = \textit{1}.	
\item \textit{diff} = \textit{1}-\textit{3} = \textit{-2}.
\item \textit{sum} = \textit{473}$\times$\textit{$10^{-2}$} = \textit{4}.
 	
\item The record \textit{(root, 4)} is stored in \textit{TAT}, because no other \textit{sum} is present in \textit{TAT} with a value of \textit{4}. If another entry was indeed present in \textit{TAT} with \textit{sum} = \textit{4}, then the entry corresponding to \textit{``root"} stored in \textit{TAT} would have been \textit{(``root", 5)}. The Tag Table (\textit{TAT}) is created simultaneously with \textit{STBE} as shown in Figure 10. The  \textit{TAT} stores the agreed upon integers corresponding to each \textit{non-variable} part of the XML.
\end{enumerate}

\begin{figure}[h!]   
\centering 
\includegraphics[width=1.5in,height=1.0in]{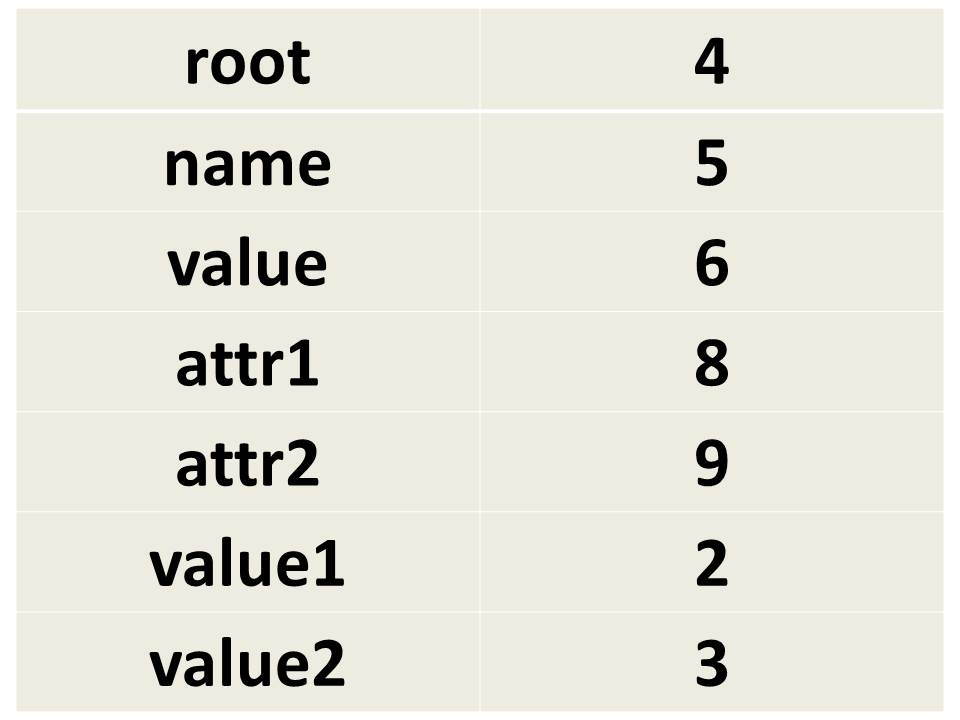}\\
\caption{\small \sl Tag Table (TAT)}
\label{fig1}
\end{figure}

\par There is provision that in the future new \textit{non-variable} parts be included in the conversation, and so \textit{noOfDigitsForNonVars} may increase. Consequently the Tag Table (\textit{TAT}) would also need to be updated in future to accommodate new \textit{non-variable} parts. In general, a RESTful communication does not need more than \textit{100} different \textit{non-variable} parts (although the same \textit{non-variable} parts may be repeated more than \textit{100} times in a message). It is therefore rare for \textit{noOfDigitsForNonVars} to be more than \textit{two}.   

\par Let us suppose that three new tags are introduced later in ``XML 1" as \textit{t1}, \textit{t2} and \textit{t3}, then the following calculations would need to done:
\begin{enumerate}
\item The tag \textit{``t1"} consists of two characters \textit{`t'} and \textit{`1'} with \textit{ST} mapping of \textit{`104'} and \textit{`340'} respectively. 
\\ \textit{sum} =  \textit{104} + \textit{340} = \textit{444}.
\item \textit{noOfNonVars} = number of new \textit{non-variable} parts introduced in the  current message (\textit{3}) + number of entries already present in the \textit{TAT} (\textit{7}) = \textit{10}.
\item \textit{noOfDigitsForNonVars} = $\ceil{\log _{10} \left( 10+1 \right)}$ = \textit{2}.	
\item \textit{diff} = \textit{2}-\textit{3} = \textit{-1}.
\item \textit{sum} = \textit{444}$\times$\textit{$10^{-1}$} = \textit{44}.
\end{enumerate}
 The newly updated Tag Table (\textit{TAT}) is shown in Fig. 11.

\begin{figure}[h!]   
\centering 
\includegraphics[width=2.0in,height=1.5in]{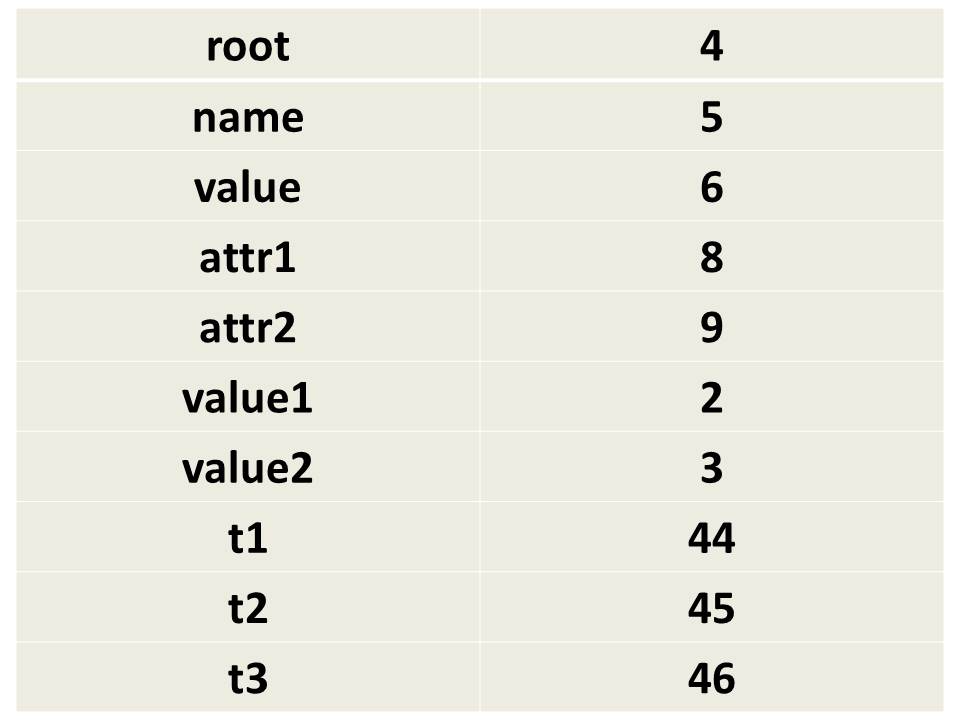}\\
\caption{\small \sl Updated Tag Table}
\label{fig1}
\end{figure}

	\item The server sends the \textit{STEnc} to the client within an HTTP response body in the form of plain-text. 
\end{enumerate}

\subsection{Symbol Table Based Decryption (STBD) and TAT Creation at Client Side}
\par Prior to receiving a response to its first request, the client does not have the Tag Table (\textit{TAT}). The client receives the \textit{STEnc} as plain-text as a part of the response. The client decrypts the \textit{STEnc} and simultaneously creates the \textit{TAT}. Each space separated sub-string of numbers in \textit{STEnc} is called a word. \textit{``(1,)"}, \textit{``0117126126104"}, \textit{``00153104104117340"} etc. are the various words in \textit{STEnc}. There are a few global variables as well. Global variables \textit{noOfNonVars} and \textit{noOfDigitsForNonVars} are initially \textit{0} and the global variable  \textit{closed} is initially \textit{1}. A global string \textit{STDec} is defined which is initially \textit{null}, and a global stack \textit{s} is defined which is initially empty. The following steps are next followed:

\begin{enumerate}
	\item The first word conveys the tag number that has been encrypted using the receiver's key. In our example, the first word is \textit{(1,)} which implies that the entire encryption \textit{(STEnc)} has been carried out using the client's key. The client can therefore decrypt the whole encryption.
	\item Step-3 through step-15 are repeated from the second word onwards in \textit{STEnc}.
	\item If the word starts with a \textit{0}, it implies a tag. The preceding \textit{0} is removed from the word.
	\\For example, to decrypt the word \textit{0117126126104}, the preceding \textit{0} is first removed and the word becomes \textit{117126126104}.
	\item If the word starts with a \textit{00}, it implies an attribute-name. The preceding \textit{00} are removed from the word.
	\\For example, to decrypt the word \textit{00153104104117340}, the preceding \textit{00} are first removed and the word becomes \textit{153104104117340}.
	\item If the word starts with a \textit{000}, it is an attribute-value. The preceding \textit{000} are removed from the word.
	\\For example, to decrypt the word \textit{000850153137820146340},  the preceding \textit{000} are removed and the word becomes \textit{850153137820146340}. 
	\item If the word does not start with a \textit{0}, it is a \textit{variable} part of the message.
	\\For example, the word \textit{122122104122} does not have a preceding \textit{0} and hence it is a \textit{variable} part of the message.	
	\item The remaining word is sliced into sub-strings of  \textit{key[8]} characters each.
	\\For example, the remaining word \textit{117126126104} is sliced into substrings of three characters each because \textit{key[8]} in this case is \textit{3}. Post slicing the client gets the following four sub-strings: \textit{``117"}, \textit{``126"}, \textit{``126"} and \textit{``104"}. 
	\item The sub-strings are next converted to integers. 
	\item If the word is \textit{non-variable} then all its integers are added and the result is stored in a variable called \textit{sum}.
	\\For the word \textit{117126126104}, \textit{sum} = \textit{117} + \textit{126} + \textit{126} + \textit{104} = \textit{473}.
     \item If the word is \textit{non-variable} then the following calculations are done:
\begin{enumerate}
\item \textit{noOfNonVars} can be calculated as a summation of the number of new \textit{non-variable} parts introduced in the current message and the number of entries already present in the \textit{TAT}.
\item \textit{noOfDigitsForNonVars}= $\ceil{\log _{10} \left( noOfNonVars+1 \right)}$	
\item \textit{diff} = \textit{noOfDigitsForNonVars}-\textit{(no. of digits in sum)}
\item \textit{sum} = \textit{sum}$\times$\textit{$10^{diff}$}
\end{enumerate} 
\par For example the following calculations are done for the word \textit{117126126104} in \textit{STEnc}: 
\begin{enumerate}
\item \textit{noOfNonVars} = number of new \textit{non-variable} parts introduced in the current message (\textit{7}) + number of entries already present in the \textit{TAT} (\textit{0}) = \textit{7}.
\item \textit{noOfDigitsForNonVars} = $\ceil{\log _{10} \left( 7+1 \right)}$ = \textit{1}.	
\item \textit{diff} = \textit{1}-\textit{3} = \textit{-2}.
\item \textit{sum} = \textit{473}$\times$\textit{$10^{-2}$} = \textit{4}.
\end{enumerate}

\par \textit{noOfDigitsForNonVars} gets calculated for each word, and gets updated when the number of \textit{non-variable} words gets increased. In general a RESTful communication does not need more than \textit{100} different \textit{non-variable} parts (although same \textit{non-variable} parts may be repeated more than \textit{100} times in a message). It is rare therefore, for \textit{noOfDigitsForNonVars} to be more than \textit{two}. This type of example was also discussed in the last sub-section.

\item The corresponding character for each number found after slicing the word is identified in \textit{ST}.
	\\In the word \textit{117126126104} for example, \textit{``117"} belongs to \textit{``r"}, \textit{``126"} belongs to \textit{``o"} and \textit{``104"} belongs to \textit{``t"}.
\item	These characters are concatenated in the same order that the corresponding integers appeared in the encrypted word. The concatenated string is stored in the variable \textit{var}.  
	\\For example, in the encrypted word \textit{117126126104} the characters \textit{``r"}, \textit{``o"}, \textit{``o"} and \textit{``t"} are concatenated to make it \textit{``root"}. \textit{var} = \textit{``root"} 		
	
	\item If \textit{var} is a \textit{non-variable} part then the following is done: 
	\begin{enumerate}
		\item The record \textit{(var, sum)} is stored in the \textit{TAT}. However if the corresponding \textit{sum} is already present in \textit{TAT} then repeat:  \textit{sum} = (\textit{sum}+\textit{1})\% \textit{$10^{noOfDigitForNonVars}$} until the \textit{sum} gets a unique value other than \textit{0}.
		\\For example the record \textit{(root, 4)} is stored in \textit{TAT}, because no other \textit{sum} is present in \textit{TAT} with a value of \textit{4}. If an entry was present in \textit{TAT} with \textit{sum} = \textit{4}, then the entry corresponding to \textit{``root"} that would be stored in \textit{TAT} would be \textit{(``root", 5)}. The Tag Table (\textit{TAT}) is created simultaneously with \textit{STBD}, which is the same as the server as shown in Figure 10.
		\item If the variable \textit{var} is a \textit{tag} and the variable \textit{closed} = \textit{0}, then \textit{STDec} = \textit{STDec} +$>$+$<$+\textit{var}. If \textit{closed} = \textit{1} then \textit{STDec} = \textit{STDec} +$<$+\textit{var}. \textit{var} is pushed into the stack \textit{s}. The variable \textit{closed} is updated to \textit{closed} = \textit{0}.
	\\For \textit{var} = \textit{``root"}, \textit{STDec} = \textit{``$<root"$}. \textit{``root"} is pushed into the stack and the variable \textit{closed} is updated to \textit{closed} = \textit{0}.
	\item If the \textit{var} is an attribute-name then \\ \textit{STDec} = \textit{STDec}+\textit{[space]}+\textit{var}+``="+```". 
	\\After making \textit{var} = \textit{``attr1"}, \textit{STDec} = \textit{``$<$root attr1=`"}.
	\item If the \textit{var} is an attribute-value then \textit{STDec} = \textit{STDec}+\textit{var}+"'".
	\\After making \textit{var} = \textit{``value1"}, \textit{STDec} = \textit{``$<$root attr1=`value1'"}.
	\end{enumerate}
	
	\item If \textit{var} is a \textit{variable} part then the steps below are followed: 
	\begin{enumerate}
	 \item If the variable \textit{closed} = \textit{0} then \textit{STDec} = \textit{STDec}+$>$+\textit{var}. If \textit{closed} = \textit{1} then \textit{STDec} = \textit{STDec}+\textit{var}. The variable \textit{closed} is updated to \textit{closed} = \textit{1}.
 	\item Based on the  data type of the decrypted \textit{variable} part, the client decides on the data type contained by its container tag. For example the \textit{variable} part \textit{``iiti"} indicates that the container tag \textit{$<$name$>$} contains a string data type.
	\end{enumerate}
	\item If the word comprises a single character 0, a `pop' operation is done on the stack \textit{s} to find the innermost opened tag. If \textit{closed} = \textit{0} then
 		 \textit{STDec} = \textit{STDec}+$>$+$<$/+pop(\textit{s})+$>$. If \textit{closed} = \textit{1} then
 		 \textit{STDec} = \textit{STDec}+$<$/+pop(\textit{s})+$>$. Update \textit{closed} = \textit{1}.
\end{enumerate}

\par Subsequent to working on all the words of \textit{STEnc}, the client gets \textit{STDec} as:
\\ \textit{STDec} = \textit{``$<$root attr1=`value1' \\ attr2=`value2'$><$name$>$iiti$<$/name$><$value$>$2$<$/value$>$\\$<$/root$>$"}. 
\\ The created \textit{TAT} is the same as that of the server and is shown in Figure 10. 

\subsection{Tag Table Based Encryprion (TATBE)}

\par The \textit{TATBE} exists at both the client and server sides. After the creation of the Tag Table (\textit{TAT}) both the client and server use it to encrypt further communications. In between a conversation, if the server introduces a new \textit{non-variable} part then the new \textit{non-variable} part gets encrypted using Symbol Table Based Encryption (\textit{STBE}) and the rest of the \textit{non-variable} parts gets encrypted using \textit{TATBE}. A client never needs to introduce a new \textit{non-variable} part of the message. The \textit{variable} parts of the message always gets encrypted using \textit{STBE}. 

\textit{TATBE} is very similar to \textit{STBE} with the only difference being that the Symbol Table (\textit{ST}) is used for character-by-character encryption of the message, and here we use the Tag Table (\textit{TAT}) to map the \textit{non-variable} parts with unique numbers.

\par Lets assume that the encrypted message gets stored in a string type global variable \textit{TATEnc}. The initial value of \textit{TATEnc} is \textit{null}. The \textit{non-variable} and \textit{variable} parts of the XML are considered in the same order in which they appear in the XML. The \textit{non-variable} and \textit{variable} parts of the XML together constitute a \textit{word}. For example, in \textbf{XML1} the order of occurrence of the words is \textit{``$<$root$>$"}, \textit{``attr1"}, \textit{``value1"}, \textit{``attr2"}, \textit{``value2"}, \textit{``$<$name$>$"}, \textit{"iiti"}, \textit{``$<$/name$>$"}, \textit{``$<$value$>$"}, \textit{``2"}, \textit{``$<$/value$>$"} and \textit{``$<$/root$>$"}.

\par Just like \textit{STBE} there is a comma separated list of tag numbers appended at the beginning of \textit{TATEnc} followed by a space. This list of tag numbers is an indication to the receiver on which tag and its children have been encrypted using the client's key. 
\\For example, the number \textit{(1,)} at the beginning of \textit{TATEnc} implies that the receiver has access to the tag \textit{``root"} and its children in \textbf{XML1} (i.e the whole document). The number \textit{(2,)} at the beginning of \textit{TATEnc} implies that the receiver has access to the tag \textit{``name"} and its children in \textbf{XML1}. The presence of \textit{(2,3,)} at the beginning of \textit{TATEnc} indicates that the receiver has access to the tags \textit{``name"} and \textit{``value"} and their children in \textbf{XML1}. We assume here that the whole XML has been encrypted using the client's key. \textit{TATEnc} is updated to:  \textit{TATEnc} = \textit{TATEnc} + \textit{``1,"}. The steps involved in the \textit{TATBE} are as follows.

\begin{enumerate}
	\item Step-2 through step-7 are repeated for each word in the XML. 
	\item If the word is a tag, the \textit{TAT} is searched for this word. If the word is found in the \textit{TAT}, its corresponding integer is fetched. The fetched integer is converted into a string and is appended with a \textit{0}. We store this string of numbers in variable the \textit{var}. \textit{TATEnc} is next updated to: \textit{TATEnc} = \textit{TATEnc} + \textit{[space]}+\textit{var}. If the tag is not available in \textit{TAT} then go to step-6.
	\\For example, the first word of \textbf{XML1} is a tag \textit{``root"}. This tag is found in \textit{TAT}. The corresponding integer of the tag, \textit{4}, is fetched  from the \textit{TAT}. This integer is converted to a string and appending with a \textit{0}. \textit{var} = \textit{04}. \textit{TATEnc} is updated to: \textit{TATEnc} = \textit{TATEnc}+\textit{[space]}+\textit{``04"}. 
	\item If the word is an attribute-name, the \textit{TAT} is searched for this word.  If the word is found in the \textit{TAT}, its corresponding integer is fetched. The fetched integer is converted into a string and is appended with with a \textit{00}. This string of number is stored in the variable \textit{var}. \textit{TATEnc} is next updated as: \textit{TATEnc} = \textit{TATEnc} + \textit{[space]}+\textit{var}. If the attribute-name is not available in \textit{TAT} then goto step-6.
	\\For example, the second word of \textbf{XML1} is the attribute-name \textit{``attr1"}. This attribute-name is available in \textit{TAT}. The corresponding integer of the attribute-name, \textit{8}, is fetched  from the \textit{TAT}. This integer is converted to a string and appended with \textit{00}. \textit{var} = \textit{008}. \textit{TATEnc} is updated to: \textit{TATEnc} = \textit{TATEnc}+\textit{[space]}+\textit{``008"}. 
	\item If the word is an attribute-value, the \textit{TAT} is searched for this word.  If the word is found in the \textit{TAT}, its corresponding integer is fetched. The fetched integer is converted into a string and is appended with a \textit{000}. This string of numbers is referred to as \textit{var}. \textit{TATEnc} is next updated as: \textit{TATEnc} = \textit{TATEnc} + \textit{[space]}+\textit{var}. If the attribute-name is not available in \textit{TAT} then goto step-6. 
\\For example, the second word of \textbf{XML1} is the attribute-value \textit{``value1"}. This attribute-name is available in \textit{TAT}. The corresponding integer of the attribute-name, \textit{2}, is fetched  from the \textit{TAT}. This integer is converted to a string and appending with \textit{000}. \textit{var} = \textit{0002}. \textit{TATEnc} is updated to: \textit{TATEnc} = \textit{TATEnc}+\textit{[space]}+\textit{``0002"}. 

		\item \textit{0} is used to represent the closing tag. Whenever a closing tag occurs, \textit{0} is concatenated with the string \textit{TATEnc} separated by a space.
		\item If the word constitutes the \textit{non-variable} part of the XML which is not found in \textit{TAT}, the word is converted into a string of numbers using \textit{STBE}. \textit{TATEnc} is updated as \textit{TATEnc} = \textit{TATEnc} + \textit{[space]} + \textit{result of STBE for word}.	
		\item \textit{STBE} is explored for the given \textit{variable} word and \textit{TATEnc} is updated as: \textit{TATEnc} = \textit{TATEnc} + \textit{[space]} + \textit{result of STBE for the word}.
\end{enumerate}

On completion of the encryption of \textbf{XML 1} the encrypted message \textit{TATEnc} looks like  \textit{TATEnc} = \textit{``1, 04 008 0002 009 0003 05 122122104122 0 06 349 0 0"}.
\par As long as an entry for a word is present in \textit{TAT}, the server does not use \textit{STBE} to encrypt the word. If, however, a new \textit{non-variable} part  is introduced in the XML as shown in \textbf{XML 2}, the newly introduced \textit{non-variable} part would need to be encrypted using \textit{STBE} while the rest of the \textit{non-variable} parts would  be encrypted using \textit{TATBE}. For the XML given in \textbf{XML 2} the final encrypted message becomes:
\textit{TATEnc} = \textit{``1, 04 008 0002 009 0003 05 122122104122 0 06 349 0 0116850 153340 0 0"}.
\\
\\
\textbf{XML 2:}
\begin{verbatim}
<root attr1="value1" attr2="value2">
<name>iiti</name>
<value>2</value>
<nv>a1</nv>
</root>
\end{verbatim}

\subsection{Tag Table Based Decryption (TATBD)}

\par \textit{TATBD} stands for \textit{TAT} based decryption and this exists at both the client and server ends. The Client gets \textit{TATEnc} as plain-text in the response whereas the server gets it within the request body. The receiver decrypts the \textit{TATEnc}. Each space separated sub-string of numbers in the \textit{TATEnc} is called a word. \textit{(1,)}, \textit{04}, \textit{008} etc. are the various words in \textit{TATEnc}. We define a global variable called \textit{closed} that is initially \textit{1}, a globally defined string \textit{TATDec} that is initially \textit{null}, and a global stack \textit{s} that is initially empty. 

\begin{enumerate}
	\item The first word indicates the tag numbers that have been encrypted using the receiver's key. In our running example the first word is \textit{(1,)}. This indicates that the whole encryption \textit{(TATEnc)} has been done using the client's key, and therefore the client can decrypt the whole XML.
	\item Step-3 through 12 are repeated for the second word onwards in \textit{TATEnc}.
	\item If the word starts with a \textit{0}, it implies that it is a tag. The preceding \textit{0} is removed from the word and it is converted to an integer.
	\\For example to decrypt the word \textit{04},  the preceding \textit{0} is first removed and it becomes \textit{4}. The same is then converted to an integer.
	\item If the word starts with \textit{00}, it implies that it is an attribute-name. The preceding \textit{00} is removed from the word and the same is converted to  an integer.
	\\For example to decrypt the word \textit{008}, the preceding \textit{00} is first removed and it becomes \textit{8}. The same is then converted to an integer.
	\item If the word starts with \textit{000}, it implies that it is an attribute-value. The preceding \textit{000} is removed from the word and the same is converted to an integer.
	\\For example to decrypt the word \textit{0002}, the preceding \textit{000} is removed and it becomes \textit{2}. The same is converted to an integer.
	\item If the word does not have a preceding \textit{0}, it implies that it is the \textit{variable} part of the message. The variable part of the message gets decrypted using the \textit{STBD}.	
	\item The \textit{non-variable} words are first searched in \textit{TAT}. If the word is not found in \textit{TAT} then the same is decrypted using the \textit{STBD}. If, however, the word is found in the \textit{TAT} then the corresponding name is fetched and stored in the variable \textbf{var}.  	
	\item If the word is a \textit{tag} and \textit{closed} = \textit{0} then \textit{TATDec} = \textit{TATDec} +$>$+$<$+\textit{var}. If \textit{closed} = \textit{1} then \textit{TATDec} = \textit{TATDec} +$<$+\textit{var}. \textit{var} is pushed into the stack \textit{s} and  \textit{closed} is updated: \textit{closed} = \textit{0}.
	\\For \textit{var} = \textit{``root"}, \textit{TATDec} = \textit{``$<$root"}. Word \textit{``root"} is pushed into the stack and  \textit{closed} is update \textit{closed} = \textit{0}.
	\item If the word is an attribute-name then \\ \textit{TATDec} = \textit{TATDec}+\textit{[space]}+\textit{var}+``="+```". 
	\\After making \textit{var} = \textit{``attr1"}, \textit{TATDec} = \textit{``$<$root attr1=`"}.
	\item If the word is an attribute-value then \textit{TATDec} = \textit{TATDec}+\textit{var}+``'".
	\\After making \textit{var} = \textit{``value1"}, \textit{TATDec} = \textit{``$<$root attr1=`value1'"}.
	\item If the word is a \textit{variable} part of the XML and \textit{closed} = \textit{0} then \textit{TATDec} = \textit{TATDec}+$>$+\textit{var}. If \textit{closed} = \textit{1} then \textit{TATDec} = \textit{TATDec}+\textit{var}. Variable \textit{closed} is update to \textit{closed} = \textit{1}.  
	\item If a single character \textit{0} is found as a word then a pop operation is done on the stack \textit{s} to find innermost opened tag. If \textit{closed} = \textit{0} then
 		 \textit{TATDec} = \textit{TATDec}+$>$+$<$/+pop(\textit{s})+$>$. If \textit{closed} = \textit{1} then
 		 \textit{TATDec} = \textit{TATDec}+$<$/+pop(\textit{s})+$>$. \textit{closed} is updated: \textit{closed} = \textit{1}. 
\end{enumerate}

Subsequent to working on all the words of \textit{TATEnc} the client gets \textit{TATDec} as \textit{``$<$root attr1=``value1" \\ attr2=``value2"$><$name$>$iiti$<$/name$><$value$>$2\\$<$/value$><$/root$>$"}.  

\subsection{Communication in web service Composition Scenario}

\par In a web service composition scenario, there is multiple service providers that work on the same message but need to have access to only some parts of the message. To realize this, the proposed approach encrypts different parts of the message with different keys. Lets consider a scenario where there is one main server (Service Provider) S and two other service providers SP1 and SP2. They commonly agree on a \textit{key} called \textit{group\_key}. 

\par A client C sends a request to the main server S. S needs to reply to the request with XML 2. In putting together the reply, S requires the services of SP1 and SP2. SP1 is required to update the value between the \textit{$<$name$>$} tag and SP2 is required to update the values between the \textit{$<$value$>$} and \textit{$<$nv$>$} tags. SP1 and SP2 must not be able to access or update any other tags that do not belong to them. In this situation, there are three symmetric keys in use. The first one is the \textit{key} between S and SP1 (say K1), the second one is the \textit{key} between S and SP2 (say K2) and the third one is the agreed upon \textit{group key} between all S, SP1 and SP2 (Say K3). The common tags of the XML will be encrypted using the  \textit{group\_key} K3. Tags that are only supposed to be updated by S and SP1 must be encrypted using K1 and tags that are only supposed to be updated by S and SP2 must be encrypted using K2.   

\par The first word of the encrypted message tells SP1 and SP2 about one or more tag numbers separated by a comma that they are supposed to process. If secret key of SP1-S is \textit{[12, 6, 1, 1, 1, 14, 4, 1, 3, 2]}, secret key of SP2-S is \textit{[6, 12, 1, 0, 1, 14, 3, 1, 3, 2]} and \textit{group\_key} is \textit{[7, 10, 0, 0, 1, 14, 3, 0, 3, 2]}. SP1 is supposed to process the second tag i.e \textit{name} and its child. SP2 is supposed to process the third and fourth tags i.e \textit{value} and \textit{nv} and their children.  \textit{TAT} based Message from S to SP1 is: \textit{``2, 01 009 0002 003 0004 05 122122104122 0 07 313 0 08 356290 0 0"}. \textit{ST} based Message from S to SP1 is: \textit{``2, 0232325325180 00137180180232257 000136137126157314257 00137180180232226 000136137126157314226 0116153109146 122122104122 0 0291356265326320 313 0 0410291 356290 0 0"}. \textit{TAT} based Message from S to SP2 is: \textit{``3,4, 01 009 0002 003 0004 05 122122104122 0 07 313 0 08 356290 0 0"}. \textit{ST} based Message from S to SP2 is: \textit{``3,4, 0232325325180 00137180180232257 000136137126157314257 00137180180232226 \\ 000136137126157314226 0116153109146 122122104122 0 0291356265326320 313 0 0410291 356265 0 0"}.

\par This web service composition scenario is just an example. Various other compositions are also possible.

\subsection{Message Authentication in Web Service Composition}
\par Message authentication is a compulsory step in any conversation where there is a possibility of updates by others in the middle because encrypted message can also be changed. For message authentication, we use existing algorithms like MD5, SHA1 etc. To exemplify this, we know from the earlier example that the tag that belongs to S and SP1 is \textit{$<$name$>$}. As both S and SP1 want a confirmation that tag \textit{$<$name$>$} which is the second tag is not changed by an intermediary. Therefore, to achieve message authentication after encrypting the tag \textit{$<$name$>$}, the sender (S) prefixes K1 to the encrypted tag and creates a hash (say MD5) of it . The overall hash is attached at the end of the corresponding tag.

\par The sequence of steps followed by the sender (S) and receiver (SP1) is:
\begin{enumerate}
\item Sender encrypts the message using \textit{TATBE} as above.
\item The \textit{TAT} based encryption of the \textit{$<$name$>$} tag is \textit{``05 122122104122 0"}. This is appended with the  private key K1 by the sender as \\ \textit{``[12,6,1,1,1,14,4,1,3,2]05 122122104122 0"}.
\item The sender calculates the MD5 of \textit{``[12,6,1,1,1,14,4,1,3,2]05 122122104122 0"} as \textit{``adc1aeffe1fe867740f976fd55c0c481"} (say \textit{D1}).
\item The \textit{TAT} based encryption of the \textit{$<$root$>$} tag is \textit{``01 009 0002 003 0004 05 122122104122 0 07 313 0 08 356290 0 0"}. This is appended with the  group key K3 by the sender as \textit{``[7,10,0,0,1,14,3,0,3,2]01 009 0002 003 0004 05 122122104122 0 07 313 0 08 356290 0 0"}.
\item The sender (S) calculates the MD5 of \textit{``[7,10,0,0,1,14,3,0,3,2]01 009 0002 003 0004 05 122122104122 0 07 313 0 08 356290 0 0"} as\\ \textit{``72afa9838090da9c5d82d2060c42f48c"} (say \textit{D2}).
\item Before sending the reply the sender appends these digests just after closing the corresponding tag for which digests have been calculated.
\textit{D1} is appended after closing the second tag and \textit{D2} is appended after closing of the \textit{``$<$root$>$"} tag. 
\item The sender sends the message: \textit{``2, 01 009 0002 003 0004 05 122122104122 0 D1 07 313 0 08 356290 0 0 D2"}. There is no need to authenticate the third and fourth tags because these do not belong to SP1.
\item The first word \textit{``(2,)"} conveys to the receiver that it has to work on the second tag. It is known to all that the first tag \textit{``$<$root$>$"} is encrypted using a group key, as this tag contains all other tags.
\item The \textit{variable} part at the end of a tag that contains alphabetic characters is the message digest of the corresponding tag. For example, D1 is present just after closing the second tag, which is why the receiver considers it as a digest of the second tag and its children. Similarly, D2 is the digest of the first tag (\textit{``$<$root$>$"}) and its children. Although the whole \textit{``$<$root$>$"} tag and its children are not encrypted using the client's key, the overall digest D2 is created by appending the K3.
\item The receiver takes out the second word of the encrypted message and D1, and further calculates its digest in the same way as the sender did in step-3 using K1. If the digest calculated by the receiver is the same as D1 then the message is authentic. If they are not the same, the message will be rejected.  
\item The receiver calculates the digest of the whole message that the sender did it in step-5 using K3. If the digest calculated by the Receiver is the same as D2 then the message is authentic. If they are not same, the message will be rejected.
\end{enumerate}

\par A Similar process is followed to authenticate the Symbol Table Based Encryption. In case of communication between S and SP2 the sender will use K2 instead of K1. The message from S to SP2 is:  \textit{``2, 01 009 0002 003 0004 05 122122104122 0  07 313 0 5f4ffdd89acc919420ac885e6017bcfc 08 356290 0 44e9e15af5f4289bab86c90e0d9398d1 0 72afa9838090da9c5d82d2060c42f48c"}.

\subsection{Working With JSON Data}
\par JSON has a different structure from XML but both are interchangeable with each other. We can easily convert XML 1 to JSON 1 as shown. 

\par We treat JSON names, values of a JSON name that start with a \textit{`-'} as \textit{non-variable} words and double quoted JSON names as \textit{variable} words. An array in JSON is considered a repetition of a tag and the values in between them are consecutive elements of an array. Closing the curly bracket denotes the end of the innermost open tag.
\par As XML and JSON are equivalent, their encryption as string of numbers must also be the same. 

\begin{table*}[]
\centering
\caption{Size comparison of various encryption techniques on various XML}
\label{my-label}
\begin{tabular}{lccccccccc}
\hline
\multicolumn{1}{c}{\textbf{Column 1}} & \textbf{Column 2} & \textbf{Column 3} & \textbf{Column 4} & \textbf{Column 5} & \textbf{Column 6} & \textbf{Column 7} & \textbf{Column 8} & \textbf{Column 9} & \textbf{Column 10} \\ \hline
\multicolumn{1}{c}{12}                & 12                & 161               & 6                 & 101               & 318               & 473               & 441               & 565               & 376                \\
\multicolumn{1}{c}{18}                & 45                & 637               & 24                & 404               & 1209              & 1665              & 1625              & 2185              & 1480               \\
\multicolumn{1}{c}{22}                & 89                & 1235              & 48                & 808               & 2397              & 3245              & 3213              & 4345              & 2938               \\ 
\multicolumn{1}{c}{18}                & 45                & 627               & 24                & 627               & 1432              & 1965              & 1921              & 2845              & 2140               \\
\multicolumn{1}{c}{8}                 & 27694             & 341167            & 27644             & 64961             & 433776            & 615258            & 578380            & 637558            & 360770             \\
\multicolumn{1}{c}{1}                 & 1                 & 39                & 1                 & 1                 & 41                & 64                & 40                & 59                & 9                  \\
\multicolumn{1}{c}{3}                 & 3                 & 12                & 1                 & 12                & 26                & 89                & 57                & 57                & 51                 \\
\multicolumn{1}{c}{1}                 & 1                 & 7                 & 1                 & 17                & 25                & 32                & 40                & 59                & 57                 \\ \hline
\multicolumn{10}{l}{\scriptsize{\textbf{Column1} represents the number of unique \textit\{non-variable\} parts of the document.}}                                                                                                                  \\
\multicolumn{10}{l}{\scriptsize{\textbf{Column2} represents the number of \textit\{non-variable\} parts used in the document.}}                                                                                                        \\
\multicolumn{10}{l}{\scriptsize{\textbf{Column3} represents the number of characters in the \textit\{non-variable\} parts.}}                                                                                                           \\
\multicolumn{10}{l}{\scriptsize{\textbf{Column4} represents the number of \textit\{variable\} parts.}}                                                                                                                                 \\
\multicolumn{10}{l}{\scriptsize{\textbf{Column5} represents the number of characters in the \textit\{variable\} parts.}}                                                                                                               \\
\multicolumn{10}{l}{\scriptsize{\textbf{Column6} represents the total number of characters in the XML (Including new line characters and a few space characters).}}                                                                   \\
\multicolumn{10}{l}{\scriptsize{\textbf{Column7} represents the number of characters in AES (Rijndael 256) encryption.}}                                                                                                              \\
\multicolumn{10}{l}{\scriptsize{\textbf{Column8} represents the number of characters in 3DES encryption.}}                                                                                                                            \\
\multicolumn{10}{l}{\scriptsize{\textbf{Column9} represents the number of characters in STBE encryption.}}                                                                                                                            \\
\multicolumn{10}{l}{\scriptsize{\textbf{Column10} represents Number of characters in TATBE encryption.}}                                                                                                                           
\end{tabular}
\end{table*}

\section{Advantage of the proposed approach}
\label{S:4}
\par Here we present the salient feature of the proposed technique highlighting its strength as compared to existing techniques.

\subsection{Need for XML canonicalization is eliminated} As  mentioned earlier, canonicalization converts logically equivalent XML documents to ones with identical physical structures. We require XML canonicalization if we encrypt an XML or part of the XML byte-by-byte. Formal security techniques provided by w3c and OASIS for SOAP based web services [3] need XML canonicalization.  But since our encryption technique is a text based substitution (Character to number), we do not need XML canonicalization.

\subsection{Need for content negotiation is eliminated} As the encryption of XML and JSON that represent the same data results in identical messages in the proposed approach. The message may also be decrypted to any of the two forms. The approach therefore seamlessly works with both XML and JSON contents. 

\subsection{No need to send extra HTTP headers} We do not send any extra HTTP headers as in the case of Serme, G. et. al [11]. In our case the agreement on a key between a service provider and its client results in a compromise of statelessness of RESTful web services to a very small extent.

\subsection{Resource representation in encrypted form} In the dynamic web services composition where a service provider may enter or leave the group randomly, the resource itself becomes a sensitive document. Putting the resource representation in the public domain is not good for such a scenario. As already discussed in our approach, the client requests a resource whose representation is not known to it with a GET method and the encrypted resource is transferred using Symbol Table Based Encryption (STBE). If the client uses an empty POST or PUT the request the server explicitly  sends the encrypted resource representation to the client using \textit{STBE}.  

\subsection{Suitable for RESTful web services composition} The proposed approach facilitates different parts of the same message to be encrypted with different keys. The various intermediaries are notified about the tag number on which they have to work. An intermediary therefore does not get any sense of others' data. If an intermediary does try to change others' data, the receiver will simply reject it because of message authentication.  

\subsection{Size of encrypted message}
The size of an HTTP message should be as small as possible in RESTful communication. The encrypted message created using this approach is much smaller in size than that of other existing algorithms. Further, if the number  of \textit{non-variable} characters is larger than the number of \textit{variable} characters, the size of the encrypted message is even smaller than the size of the actual message itself. However if the number of \textit{variable} characters dominates over the number of \textit{non-variable} (which is relatively rare) characters the size of the encrypted message using our approach may sometimes become larger than that of existing approaches. In this case the size of  the encrypted message largely depends on \textit{key[8]}. Table 1 describes the size comparison of encrypted messages using various encryption algorithms on various types of XML data. Here we do encryption on the whole XML, not on a given part of the XML. The key used for STBE and TATBE is \textit{[12,6,1,1,1,14,4,1,3,2]}. Here we include various characters like $<, >, /, ``, =$ etc used for the \textit{non-variable} parts in the count of the number of characters used in the \textit{non-variable} parts.

\subsection{Hard to attack} This is especially true in the condition that a service provider within the group is the attacker. This, in fact, makes this approach especially useful. The following points make this encryption hard to attack:
\begin{enumerate}
\item The \textit{10-element} key is generated randomly and sent to the other party using TLS. It is therefore not possible for other service providers to sniff the key.
\item A brute force attack is not possible. This is so because first, it is very hard to go through all possible values for all ten elements, and then to create various tables for all possibilities, and further to decrypt a part of the XML that does not belong to the  attacker with all possible keys. Even if the attacker manages the above, it will not be sure which one is the correct decryption.
\item Message authentication prevents  an attacker from changing the unauthorized part of the XML through hit and trial.
\item The \textit{Known plain-text attack} [12] is the most dangerous for this approach. This attack is possible if the attacker has a valid plain-text/cipher-text pair. Using this, the attacker can try to guess the further encrypted message or even the key. If somehow the attacker is able to map between the various characters and the number or the tag-name and the number, then all of the encryption will be compromised. The question here is how will the attacker get a valid cipher-text/plain-text pair in a web service composition scenario. 
\item The \textit{chosen plain-text attack}  [13] is a attack that presumes that the attacker can get a cipher-text for any random plain-text and can try to guess the further conversation or the key itself. In case of web service composition the attacker is a service provider and would have a different shared key, and therefore it can not possibly get an encrypted message with other's key.
\item Various security aspects on RESTful web services are described by OWASP [14]. Here we are dealing with only the message level security. In our approach XML and JSON input validation and message integrity are provided through message authentication of the XML or JSON document.
\end{enumerate}

\section{Conclusion}
\label{S:5}
In this paper, we proposed a novel approach for message level security in RESTful web services. This approach is a kind of substitution cipher which replaces different characters by a unique integer. Various tag-names, arguments and values of arguments were also replaced by corresponding unique integers. Substitution from character to number removes the need for XML canonicalization. This approach removes the need for content negotiation for the same resource among service providers and clients, and also reduces the size of the overall encrypted message. We applied this approach on several XML and JSON data and found that for most of the cases our approach resultd in smaller sized encrypted messages than those of existing approaches. We also demonstrated the efficiency of the proposed message in web services composition scenarios through encrypting different tags with different keys, and also maintained message authentication. The discussed algorithms were implemented in JAVA.

\section*{Acknowledgment}
\label{S:6}

The authors would like to thank the ``Department of Electronics and IT (DeitY), Government of India" for funding this project.

\bibliographystyle{model1-num-names}
\bibliography{sample.bib}



\end{document}